\documentclass[sigconf, nonacm]{acmart}

\usepackage{url}
\usepackage[ruled,lined,linesnumbered]{algorithm2e}

\begin{document}

\title{Cache-Consistent Dynamic Load Balancing for\\ Kubernetes Controllers}

\author{Shoma Ansai}
\email{ansai@inet.media.kyoto-u.ac.jp}
\affiliation{%
  \institution{Kyoto University}
  \city{Kyoto}
  \country{Japan}
}

\author{Yasuo Okabe}
\email{okabe@media.kyoto-u.ac.jp}
\affiliation{%
  \institution{Kyoto University}
  \city{Kyoto}
  \country{Japan}
}

\author{Daisuke Kotani}
\email{kotani@media.kyoto-u.ac.jp}
\affiliation{%
  \institution{Kyoto University}
  \city{Kyoto}
  \country{Japan}
}

\renewcommand{\shortauthors}{Ansai et al.}

\begin{abstract}
As Kubernetes clusters grow, the scalability of controllers can become a bottleneck for the performance of the system.
Distributing the load dynamically across multiple controller instances, however, raises the following two problems, and a controller can therefore be run only as a single instance today.
The first problem is the cost of reassignment.
A controller retrieves objects on the basis of the Labels attached to them, so in a naive design in which the assigned instance is recorded in a Label on every object, the Labels must be rewritten in proportion to the total number of objects whenever instances are added or removed.
The second problem is cache consistency.
A controller consults only its own cache when it reads an object and never refers to the actual data, so the cache has to be updated explicitly at the time of a reassignment.
Furthermore, a controller manages a cache independently for each kind of object, so cache updates have to be synchronized across the kinds of objects.
We propose a method for scaling Kubernetes controllers horizontally that combines lightweight load balancing with cache synchronization.
A two-level Hash maps objects to Virtual Nodes, records their identifiers in Labels, and assigns Virtual Nodes to instances by Consistent Hashing.
Whenever instances are added or removed, it therefore suffices to update the Label value specified when objects are retrieved, and no Label on an object has to be rewritten.
The cache is also locked until the reassignment has completed, which prevents any reference to a stale cache.
The version identifier of the data store is used to synchronize the kinds of objects with one another.
We implemented the proposed method on Kubernetes and evaluated it: the processing throughput rises with the number of instances, and the time required for reassignment remains within an acceptable range.

\end{abstract}




\maketitle

\section{Introduction}\label{sec:introduction}

Kubernetes is widely used as a platform for managing containerized
workloads and services~\cite{k8s-overview}.
Its central mechanism is the controller.
In the Kubernetes control plane, the current and the desired state of
a cluster are represented as a collection of objects; a controller
watches these objects and reconciles the current state with the
desired one.
This mechanism recovers from failures and applies configuration
changes automatically, without human intervention.

In large-scale clusters, however, the scalability of controllers has
become a pressing issue: the number of objects grows, whereas a
controller is ordinarily confined to a single active instance, and a
single instance may no longer keep pace with the processing
load~\cite{ebert2024}.

Several software projects support horizontal
scaling~\cite{fluxcd,argocd,knative,kubevela,externaldns,cert-manager,ingress-nginx,prometheus-operator,kep5866}.
In these projects, however, objects are assigned to the controller
instances in advance, so that controllers cannot be scaled out
dynamically in response to the current load.
To address this limitation, Ebert~\cite{ebert2024} proposed Controller
Sharding, which employs Consistent Hashing to distribute objects
across multiple controller instances automatically by attaching the
assigned instance to each object as a Label.
Two critical problems nevertheless remain before dynamic scaling of
controllers can be achieved.

The first problem is the cost of reassignment.
A controller watches objects and receives notifications of updates to
them, selecting the objects to watch by the metadata attached to them,
which is called a Label.
If the assigned instance is simply stored in a Label and objects are
watched on the basis of such Labels, the Labels on all objects must be
rewritten whenever instances are added or removed.
This rewriting burdens the data store and the API Server in proportion
to the number of objects, a cost that may be unacceptable in a
large-scale Kubernetes deployment.

The second problem is that cache handling demands special
consideration during reassignment.
A controller instance in Kubernetes holds a cache of the object data
kept in the data store behind the API Server.
When the instance accesses an object, it consults this cache alone and
issues no query to the API Server; the cache is updated when the API
Server notifies the instance of a change to the object.
Moreover, each instance maintains a separate cache and operates on
objects in parallel for each type of object, which is called a
resource.
Consequently, the cache update for one resource may have completed
while that for another has not, and a controller can therefore process
objects on the basis of stale data.

Several existing studies preserve consistency during reassignment by
locking the cache~\cite{decandia2007,redis-cluster,annamalai2018,lakshman2010,stoica2001,li2014,hoffmann2019,delmonte2020,gu2022}.
In a Kubernetes controller, however, the watching of objects and the
updating of the cache proceed in parallel for each resource, so that
state must be synchronized across different resources; these methods
therefore cannot be applied as they are.

In this paper, we propose a horizontal scaling method for Kubernetes
controllers that solves both problems.

For the first problem, we introduce object assignment based on a
two-level Hash.
Instead of assigning objects directly to controller instances, we
interpose the concept of Virtual Nodes as an intermediate layer.
In the first stage, objects are mapped to a fixed number of Virtual
Nodes by the remainder of their hash value; in the second, Virtual
Nodes are assigned to instances by a Consistent Hash Ring.
When instances are added or removed, only the second-stage mapping is
updated, which makes it possible to change the assignment of objects
to controllers without rewriting the Labels on the objects.
Interposing a fixed number of intermediate partitions between objects
and nodes is an approach adopted by several distributed
systems~\cite{decandia2007,redis-cluster,kreps2011,annamalai2018}.
Applying it to a controller that holds a cache, however, also requires
accounting for how the cache is updated when the assignment changes,
and it therefore cannot be adopted as it stands.

For the second problem, we introduce a cache update mechanism based on
a Barrier.
When the object assignment changes, the API Server sends an
assignment-change notification to every instance, and each instance,
upon receiving it, temporarily blocks cache access to all related
objects.
The notification carries a Revision, the version identifier of the
data store, which serves to synchronize different resources with one
another.
While access is blocked, the instance removes from its cache the
objects of the Virtual Nodes it no longer holds, and Lists only the
objects of the newly assigned Virtual Nodes and adds them to the
cache.
Releasing the block once the cache update has completed prevents
access to an incomplete cache or to stale data.

We implemented and evaluated these ideas, and showed that the
processing throughput increases with the number of instances and that
the time required for a reassignment stays within an acceptable range.

This study makes the following three contributions.

\begin{itemize}
  \item We showed that dynamic horizontal scaling can be achieved
  without rewriting objects, by means of object assignment based on a
  two-level Hash and the maintenance of cache consistency through a
  Barrier mechanism.
  \item We implemented the proposed method on Kubernetes and showed
  that it is feasible as a system that actually works.
  \item We confirmed through evaluation that the processing throughput
  improves with the number of instances and that the overhead incurred
  by a reassignment stays within an acceptable range.
\end{itemize}

The rest of this paper is organized as follows.
Section~2 defines the model of Kubernetes on which this study is
based.
Section~3 presents the problems addressed in this study.
Section~4 organizes related work.
Section~5 describes the design of the proposed method.
Section~6 explains the details of the implementation.
Section~7 carries out the evaluation, and Section~8 presents the
discussion.
Finally, Section~9 states the conclusion and future work.

\section{Kubernetes Architecture}\label{sec:kubernetes-model}

This section presents a model of the Kubernetes control plane, which
is the part of the Kubernetes architecture on which this study
focuses.
The discussion in the remainder of the paper builds upon this model.

The control plane manages a collection of entities called objects,
each of which carries a desired state supplied by the user.
It continuously monitors the actual state of every object and,
whenever that state diverges from the desired state, automatically
performs the operations required to bring the former closer to the
latter.
Through the repetition of this control cycle, the system as a whole
converges to the desired state.

An object may carry key-value metadata called Labels, which serve as
filtering criteria when objects are retrieved.
Objects are classified by type, and each type of object is termed a
resource.
Objects may also stand in a parent-child relationship, in which a
child object is created on the basis of a parent object.

The control plane comprises the following three components.

\begin{itemize}
  \item API Server
  \item Data Store
  \item Controller
\end{itemize}

\begin{figure}[t]
  \centering
  \includegraphics[width=\columnwidth]{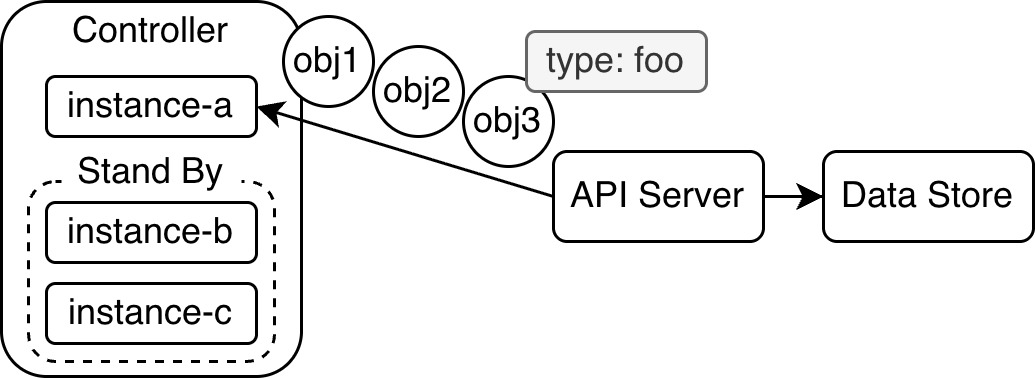}
  \Description{A block diagram showing the model of the Kubernetes control plane. A Controller block contains instance-a as active and instance-b and instance-c as Stand By. Three objects labeled obj1, obj2, obj3 of type foo point to instance-a. An API Server connects to a Data Store on the right.}
  \caption{Model of the Kubernetes control plane.}
  \label{fig:model}
\end{figure}

Figure~\ref{fig:model} illustrates the model, and each component is
described in turn below.

The API Server is the entry point through which every access to an
object passes.
Objects are created, updated, and deleted exclusively through the API
Server, and no controller ever reaches the data store directly.
When a Label is specified at retrieval time, the API Server returns
only those objects that carry the Label.

The data store holds the actual content of the objects.
It is reachable only from the API Server and is never accessed
directly by a controller instance.
The data store maintains a system-wide counter called
``\texttt{Revision}'', which is incremented on every change to the
data and thereby establishes the relative order of object states.

\begin{figure}[t]
  \centering
  \includegraphics[width=\columnwidth]{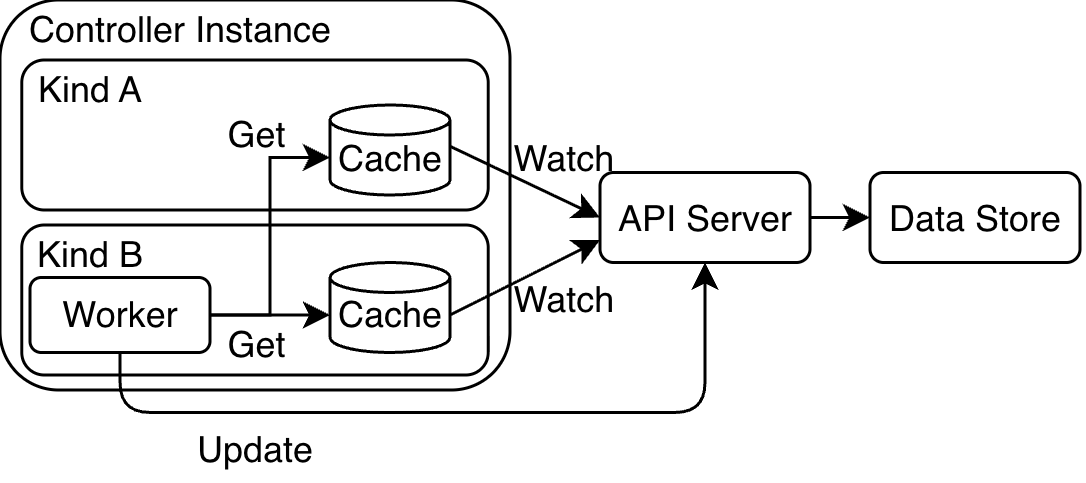}
  \Description{A block diagram showing the operation of a controller instance. The instance contains two Kind blocks, Kind A and Kind B, each with a Worker that performs Get operations on its own Cache. Each Cache receives updates via a Watch connection to the API Server. The API Server connects to a Data Store. The Worker sends Update requests to the API Server.}
  \caption{Operation of a controller.}
  \label{fig:controller}
\end{figure}

Figure~\ref{fig:controller} depicts the operation of a controller.
A controller is the entity that watches objects and performs the
operations needed to close the gap between the desired state and the
actual state; this operation is called Reconcile.
For redundancy, several processes of the same controller may be run,
and each such process is called a controller instance, abbreviated
hereafter to instance where the context permits.
Even when multiple instances are running, leader election is
ordinarily performed among them, so that only one instance operates
actively while the others remain on standby.

Every instance runs one or more workers, which carry out Reconcile.

Each instance also holds a cache that stores the latest state of the
objects for which it is responsible.
Rather than querying the API Server, an instance reads the state of an
object from this cache, which reduces the load imposed on the API
Server and the data store.

Within an instance, a separate cache and a separate set of workers
exist for each resource, so that objects are received, cached, and
processed in parallel on a per-resource basis.
Since the processing of a parent object requires the state of its
child object, a worker responsible for one resource may reference not
only the cache of its own resource but also the cache of another
resource with which it stands in a parent-child relationship.

To keep the cache of each instance current, the API Server
asynchronously notifies instances of changes to objects; this
notification mechanism is called Watch.
At start-up, an instance first retrieves the current list of the
objects for which it is responsible and stores them in its cache, an
operation called List.
Thereafter, every time the addition, modification, or deletion of an
object is notified through Watch, the instance updates its cache
accordingly.

Both the response to a List and the notifications delivered by Watch
carry the Revision in effect at that moment, and Watch is designed to
deliver, without loss, every change that occurs after that Revision.
Combining List and Watch therefore allows each instance to observe
every change to an object without loss.

\section{Problems Addressed in This Study}\label{sec:problem}

\subsection{Introduction of Horizontal Scaling}\label{subsec:problem-loadbalancing}

As described in Section~\ref{sec:kubernetes-model}, only one instance
of a controller ordinarily operates actively, even when several
instances are running.
This study, by contrast, considers making every controller instance
operate actively so as to balance the load through horizontal scaling.

Each instance is then responsible for the portion of the objects that
has been assigned to it.
The API Server notifies each instance of the range of objects for
which it is responsible, and each instance retrieves from the API
Server only those objects.
An instance is thereby kept from retrieving objects outside its
responsibility and wasting memory and network bandwidth.

Controller instances are added, stopped, and may fail dynamically.
Whenever instances are added or removed, the range of objects for
which each instance is responsible must be reconfigured; we refer to
this procedure as reassignment.

Two major problems remain to be solved before such reassignment can be
realized.

\subsection{Problem 1: State Update Load Incurred by Reassignment}\label{subsec:problem-rebalance-load}

The first is the load imposed on the control plane.
As described in Section~\ref{sec:kubernetes-model}, a controller can
obtain from the API Server only the objects that carry a given Label
by specifying that Label at retrieval time.
If the assigned controller is simply stored in a Label, however,
reassignment entails a rewrite operation proportional to the number of
affected objects.
Every such rewrite passes through the API Server and entails a write
to the data store, so that the load on the API Server and the data
store grows in proportion to the number of affected objects.
In a large-scale system, where the number of objects tends to be
enormous, this places a load on the system that cannot be ignored.

\subsection{Problem 2: Cache Updates During Reassignment}\label{subsec:problem-cache-update}

The second is the problem of cache updates.
When reassignment occurs, each instance must update its cache to match
its new set of assigned objects.
As described in Section~\ref{sec:kubernetes-model}, cache updates
proceed independently for each resource, and this parallelism gives
rise to the following three challenges.

\textbf{The Cache Consistency Challenge}\quad
When reassignment occurs, the cache update for each resource proceeds
independently, so that even after the update for one resource has
completed, the cache of another resource may still remain stale.

Because an instance references objects from its own cache, processing
that runs against a stale cache operates on outdated information.

The following scenario illustrates the problem.
Consider two resources in a parent-child relationship, both subject to
load balancing, where the parent references the child object during
its processing.

\begin{enumerate}
  \item An assignment change occurs because instances are added or
  removed.
  \item The assignment-change processing for the resource of the
  parent finishes first, and its cache is updated to reflect the new
  assignment.
  \item At this point, the assignment-change processing for the
  resource of the child has not yet finished, and the cache of the
  child remains stale.
  \item Reconcile runs for the parent and references the child object.
  \item Because the cache of the child is still stale, outdated
  content is returned, and an inconsistency arises.
\end{enumerate}

\textbf{Consistent Assignment of Parent and Child Objects}\quad
As described in Section~\ref{sec:kubernetes-model}, processing a
parent object requires referencing the state of its child object.
Were the parent and the child assigned to different instances, the
child object would be absent from the cache of the instance
responsible for the parent, and the parent could not be processed
correctly.

Objects in a parent-child relationship must therefore be assigned to
the same instance consistently, even across reassignments.

\textbf{Handling Consecutive Reassignments}\quad
The addition or removal of instances may occur in a short period of
time.
If the next notification arrives before the processing of the previous
reassignment has completed, the cache updates interleave, and cache
consistency becomes difficult to maintain.

If, for example, the completion of the first reassignment were taken
to mean that reassignment as a whole had completed, the cache would
become accessible while the second reassignment was still incomplete.

A mechanism is therefore needed that keeps the cache consistent even
when reassignments occur consecutively while a previous one is still
being processed.

\section{Related Work}\label{sec:related-works}

This section summarizes the prior work related to this study.

\subsection{Horizontal Scaling in Distributed Systems}\label{subsec:related-scaling}

As prior work on the horizontal scaling of the Kubernetes control
plane, several projects have implemented their own approaches for
practical Kubernetes controllers.

Projects such as Flux CD~\cite{fluxcd}, ExternalDNS~\cite{externaldns},
cert-manager~\cite{cert-manager}, Ingress-NGINX~\cite{ingress-nginx},
and Prometheus Operator~\cite{prometheus-operator} provide a means of
restricting the scope of each controller instance by Namespace or
Label Selector.
Flux CD, for example, lets the user attach Labels to objects manually
through the ``\texttt{--watch-label-selector}'' option, thereby
assigning each object to a particular shard.
Argo CD~\cite{argocd} shards on a per-cluster basis, each instance
being responsible for a set of managed Kubernetes clusters.
In KubeVela~\cite{kubevela}, a master instance attaches a shard-id
Label to applications through a Mutating Webhook and dispatches the
work to slave instances.
All of these approaches either require the assignment of objects to be
managed manually or depend on a specific object structure, and none of
them can therefore scale out dynamically in response to the current
load.

Knative~\cite{knative} performs Leader Election per Reconciler and per
Bucket, and distributes the work by making each instance responsible
for a subset of the Buckets.
Since every instance nevertheless watches every object, memory
consumption is not distributed, which limits the scalability gain.

KEP-5866~\cite{kep5866}, a Kubernetes Enhancement Proposal (KEP),
proposes server-side sharded LIST and WATCH.
A client specifies a hash range through the ``\texttt{shardSelector}''
parameter, and the API Server delivers only the events for the objects
that fall in that range.
The computation of the hash range and the assignment of objects are
nevertheless left to the client side, so that nothing guarantees
consistent processing across the cluster as a whole.

The prior work most directly related to this study is Ebert's
method~\cite{ebert2024}, which distributes objects across multiple
controller instances by Consistent Hashing and implements object
filtering by means of Label Selectors.
Ebert's method, however, suffers from the two problems discussed in
Section~\ref{sec:problem} when instances are added or removed, and
this study proposes a method that solves both of them.

\subsection{Two-Level Hashing in Distributed Systems}\label{subsec:loadbalance-with-two-level-hash}

The first problem, discussed in
Section~\ref{subsec:problem-rebalance-load}, stems from the design
that assigns objects directly to instances.
For problems of this kind, several distributed systems adopt a design
in which a fixed number of intermediate partitions is placed between
objects and nodes, and load balancing is carried out by a two-level
mapping~\cite{decandia2007,redis-cluster,kreps2011}.

In every one of these cases, the mapping from objects to intermediate
partitions is fixed, and only the mapping from intermediate partitions
to nodes is updated when nodes are added or removed, which keeps the
cost of reassignment low.
This design avoids the first problem of Ebert's method.
The second problem, that of cache consistency, discussed in
Section~\ref{subsec:problem-cache-update}, nevertheless arises in the
same way when a two-level Hash structure is applied.

\subsection{Data Consistency During Rescaling}\label{subsec:data-consistency-during-rescaling}

As discussed in Section~\ref{subsec:problem-cache-update}, the problem
of cache consistency arises when the Kubernetes control plane is
scaled horizontally.
Comparable problems have been treated by several existing distributed
systems that handle data consistency during rescaling.

In systems that divide data into partitions and place them on nodes,
for example, the data must be handed over to the new responsible node
when nodes are added or removed.
Mechanisms have accordingly been designed to transfer data from the
old node to the new one when partitions are
reassigned~\cite{decandia2007,redis-cluster,annamalai2018,lakshman2010,stoica2001,li2014,hoffmann2019,delmonte2020,gu2022}.

Under this design, accesses to the partition concerned are blocked
while the data is being transferred.
One might suppose that, for the Kubernetes control plane, it would
suffice to block access to the cache during reassignment.
As discussed in Section~\ref{subsec:problem-cache-update}, however,
Watch and cache updates are performed in parallel for each resource,
so that reassignment may still be in progress for one resource after
it has completed and the block has been released for another.
State must be synchronized across different resources, and existing
approaches therefore cannot be applied as they are.

In other systems, the cache is not updated at all during
rescaling~\cite{nishtala2013,nygren2010,eisenbud2016}.
There, the authoritative data is kept on a backend such as a database,
and the node to which a request is dispatched holds only a copy of it
as a cache.
When nodes are added or removed and requests are routed to a new node,
a cache miss occurs because the cache does not exist on that node, and
the correct data is obtained by re-fetching it from the backend.
Since requests are no longer routed to the old node, the problem of
referencing a stale cache does not arise either.

This approach, however, cannot be used in the Kubernetes control
plane.
As described in Section~\ref{sec:kubernetes-model}, an instance
references objects from its own cache, so that an object absent from
the cache is treated as though it did not exist.
Unless the cache is updated explicitly during reassignment, accesses
to a stale cache will therefore occur.

Existing approaches thus cannot solve the problem of cache consistency
in the Kubernetes control plane.

\section{Proposed Method}\label{sec:proposal}

This section describes the horizontal scaling method proposed in this
study.
Section~\ref{subsec:proposal-overview} gives an overview of the
proposal, and Section~\ref{subsec:watch-group} introduces the concept
of a Watch Group.
Section~\ref{subsec:two-tier-hash} then presents the design of the
two-level Hash, which forms the core of the proposal.
Section~\ref{subsec:resource-assignment} describes the management of
Virtual Node assignment in detail, and
Section~\ref{subsec:barrier-mechanism} finally explains how the
Barrier mechanism maintains cache consistency.

\subsection{Overview of the Proposal}\label{subsec:proposal-overview}

The proposed method rests on the following two ideas.

\begin{figure}[t]
  \centering
  \includegraphics[width=\columnwidth]{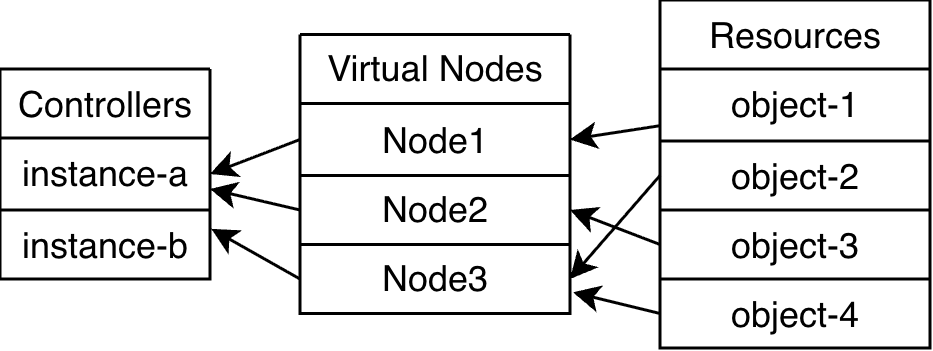}
  \Description{A diagram showing three columns: Controllers on the left with instance-a and instance-b, Virtual Nodes in the middle with Node1, Node2, and Node3, and Objects on the right with object-1 through object-4. Arrows show objects mapped to Virtual Nodes, and Virtual Nodes mapped to controller instances.}
  \caption{Object assignment by a two-level Hash.}
  \label{fig:two-level-hash}
\end{figure}

\begin{itemize}
\item Object assignment based on a two-level Hash:

Instead of assigning objects directly to controller instances, we
interpose the concept of Virtual Nodes between the two
(Figure~\ref{fig:two-level-hash}).
In the first stage, objects are mapped to a fixed number of Virtual
Nodes by the remainder of their hash value; in the second, Virtual
Nodes are assigned to controller instances by a Consistent Hash
Ring~\cite{karger1997}.
Since only the second stage is updated when instances are added or
removed, the amount of recomputation is constant and independent of
the number of objects.

\item Cache updates by a Barrier mechanism when instances are added or
removed:

When the assignment of objects changes, each instance temporarily
blocks cache access to all related objects.
The Revision provided by the data store serves to synchronize
different resources with one another.
The instance then Lists only the objects of the newly assigned Virtual
Nodes and adds them to its cache.
Releasing the block once the List has completed prevents access to an
incomplete cache.
\end{itemize}

\subsubsection{Operation Flow in the Steady State}\label{subsubsec:normal-flow}

In the steady state, the method operates as follows.
When an object is written, the API Server computes the Virtual Node
number to which the object belongs and attaches it to the object as a
Label.

Each instance retrieves from the API Server the list of the Virtual
Nodes assigned to it and, on the basis of this list, specifies a Label
so as to List and Watch only the objects that belong to the Virtual
Nodes for which it is responsible.

The instance stores the objects it receives in its own cache, updates
the cache whenever a change notification arrives, and reads from that
cache in all processing that references an object.

\subsubsection{Operation Flow During Rescaling}\label{subsubsec:reassignment-flow}

When instances are added or removed, reassignment proceeds as follows.

The API Server first detects the addition or the removal, recomputes
the Consistent Hash Ring of the two-level Hash, and notifies each
instance of its new Virtual Node assignment.
Upon receiving the notification, each instance temporarily blocks
cache access through the Barrier mechanism, Lists the objects of the
Virtual Nodes for which it has newly become responsible, and adds them
to its cache.
Once the List has completed, the instance releases the block and
returns to normal operation.

\subsection{The Watch Group Concept}\label{subsec:watch-group}

A Watch Group is a set of controller instances that share the
responsibility for reconciling objects.
The group is identified by a name called ``\texttt{watchGroup}'', and
each instance carries an identifier called ``\texttt{memberID}'' that
identifies it uniquely within the group.

If, for example, three instances named ``\texttt{worker-a}'',
``\texttt{worker-b}'', and ``\texttt{worker-c}'' belong to a Watch
Group named ``\texttt{parent-workers}'', the Parent objects as a whole
are distributed among these three instances without overlap.

In the proposed method, an instance is registered as a member of the
Watch Group when it establishes a Watch connection to the API Server,
and the registration of that instance is removed automatically when
the connection is broken.
The membership of the group is thereby kept up to date without any
explicit deregistration step.

\subsection{Object Distribution by a Two-Level Hash}\label{subsec:two-tier-hash}

As the algorithm for distributing objects across the instances of a
Watch Group, this proposal adopts Consistent Hashing~\cite{karger1997}.
Consistent Hashing has the property of minimizing the number of
objects affected when instances are added or removed, and it is widely
used in distributed caches and distributed databases, beginning with
Dynamo by DeCandia et al.~\cite{decandia2007}.

A configuration that assigns objects directly to instances by a single
Consistent Hash Ring cannot, however, be applied to the Kubernetes
control plane as it stands.
Under such a configuration, the information about the responsible
instance would be written directly into the Label of the object, so
that the Labels of every object whose owner had changed would have to
be rewritten whenever instances were added or removed.
This would reproduce Problem 1 discussed in
Section~\ref{subsec:problem-rebalance-load}.
The proposal therefore interposes Virtual Nodes as an intermediate
layer between objects and instances, and splits the Hash into two
stages.

\subsubsection{Structure of the Two-Level Hash}\label{subsubsec:two-tier-hash-structure}

The Hash is applied in two stages.
In the first stage, an object is mapped to a Virtual Node by the
remainder of its hash value modulo the fixed number $V$ of Virtual
Nodes.
$V$ is determined before the system starts operating and does not
change as instances are added or removed, so that the first-stage
mapping remains constant throughout the lifetime of the system.
In the second stage, a Consistent Hash Ring maps Virtual Nodes to
instances, and only this ring is updated when instances are added or
removed.

The instance responsible for a given object is determined by the
following equation.
\[
  \mathrm{Owner}(\mathit{object}) = \mathrm{ConsistentHash}(\mathrm{hash}(\mathit{objectKey}) \bmod V)
\]

When instances are added or removed, ownership has to be recomputed
only for the $V$ Virtual Nodes.
Since $V$ is a constant independent of the number of objects $M$, the
cost of recomputation is $O(V)$, that is, a fixed cost that does not
depend on the number of objects.

Furthermore, because the Label of an object holds its Virtual Node
number and the first-stage mapping is constant, no Label on any object
ever has to be rewritten when instances are added or removed.
This resolves Problem 1 discussed in
Section~\ref{subsec:problem-rebalance-load}.

\subsection{Management of Object Assignment}\label{subsec:resource-assignment}

This section explains how objects are assigned by means of the
two-level Hash.

\subsubsection{Filtering Mechanism}\label{subsubsec:filtering-mechanism}

When an object subject to load balancing is written, the API Server
computes the Virtual Node number to which the object belongs on the
basis of the first-stage modular hash and attaches it to the object as
a Label.
Since the first-stage mapping is constant, the Virtual Node number of
an object does not change throughout the lifetime of the object.

The API Server may be run with multiple replicas for load balancing
and high availability.
Were each replica to hold the membership state independently, the
state would diverge across replicas and inconsistencies could arise in
the assignment of objects.
In this proposal, the membership of a group is therefore stored in the
data store, so that every replica refers to the same membership state.

Each API Server replica watches the membership state in the data
store, and each instance watches, through the API Server, the list of
the Virtual Nodes for which it is responsible.
When the API Server detects a change in membership, it recomputes the
list of the Virtual Nodes assigned to each instance from the
second-stage Consistent Hash Ring and notifies every instance whose
assignment has changed of its new assignment.
The hash function used for the Consistent Hash Ring is deterministic,
that is, its output for a given input string is always the same,
regardless of the environment or the process.
This ensures that every API Server replica builds an identical Hash
Ring.

Upon receiving a notification of a change in its Virtual Node
assignment, an instance first stops the Watch currently in progress.
Since the instance keeps the previous list of Virtual Nodes as
``\texttt{past\allowbreak Virtual\allowbreak Nodes}'', it can compare that list with the
new assignment and List the newly assigned objects.
Once the List has completed, it resumes the Watch with a Label based
on the list of Virtual Nodes it has received.
Algorithm~\ref{alg:vn-assignment-change} presents this procedure as
pseudocode.
If another assignment-change notification arrives before the List has
completed, the instance waits for the current List to complete and for
``\texttt{past\allowbreak Virtual\allowbreak Nodes}'' to be updated before it Lists again.
Events are thereby handled correctly even when the addition or removal
of instances occurs in a short period of time.

To prevent Reconcile from running on objects that remain in the
WorkQueue but are no longer the responsibility of the instance, each
instance checks before running Reconcile whether the Virtual Node
associated with the object is assigned to itself, and skips Reconcile
if it is not.
Algorithm~\ref{alg:vn-filter} presents this check.

\begin{algorithm}[t]
\caption{Instance processing on Virtual Node assignment change}
\label{alg:vn-assignment-change}
\SetKwFunction{StopWatch}{StopWatch}
\SetKwFunction{ListResources}{List}
\SetKwFunction{StartWatch}{StartWatch}
\SetKwProg{Fn}{Function}{:}{}
\KwIn{$\mathit{newVNs}$: list of newly assigned Virtual Nodes, \\
      $\mathit{revision}$: Revision of the assignment change}

\Fn{\textnormal{OnVNAssignmentChange}($\mathit{newVNs}$, $\mathit{revision}$)}{
  \StopWatch{}\;
  \BlankLine
  $\mathit{addedVNs} \leftarrow \mathit{newVNs} \setminus \mathit{pastVirtualNodes}$\;
  \BlankLine
  \If{$\mathit{addedVNs} \neq \emptyset$}{
    $\mathit{labelFilter} \leftarrow$ \texttt{"vn $\in$ addedVNs"}\;
    \ListResources{$\mathit{labelFilter}$}\tcp*{List only objects of newly added VNs}
  }
  \BlankLine
  $\mathit{pastVirtualNodes} \leftarrow \mathit{newVNs}$\;
  $\mathit{labelFilter} \leftarrow$ \texttt{"vn $\in$ newVNs"}\;
  \StartWatch{$\mathit{labelFilter}$}\;
}
\end{algorithm}

\begin{algorithm}[t]
\caption{Virtual Node assignment check before Reconcile}
\label{alg:vn-filter}
\SetKwFunction{ProcessItem}{ProcessItem}
\SetKwFunction{VNLabelOf}{VNLabelOf}
\SetKwFunction{Reconcile}{Reconcile}
\SetKwProg{Fn}{Function}{:}{}

\Fn{\ProcessItem{$\mathit{object}$}}{
  $\mathit{vn} \leftarrow$ \VNLabelOf{$\mathit{object}$}\;
  \If{$\mathit{vn} \notin \mathit{assignedVNs}$}{
    \Return
  }
  \Reconcile{$\mathit{object}$}\;
}
\end{algorithm}

\subsubsection{Assigning Parent and Child Objects to the Same Instance}\label{subsubsec:parent-child-same-pod}

As described in Section~\ref{subsec:problem-cache-update}, objects in
a parent-child relationship must be assigned to the same instance.

To solve this problem, the proposal provides a mechanism by which the
value passed to the first-stage hash function, the Hash Key, can be
specified through the Label of the object.
Setting this value identically on objects that stand in a parent-child
relationship maps them to the same Virtual Node in the first stage and
thereby assigns them to the same instance.

\subsection{Maintaining Cache Consistency through the Barrier Mechanism}\label{subsec:barrier-mechanism}

This section describes the Barrier mechanism, which solves the
challenge of updating the cache at reassignment discussed in
Section~\ref{subsec:problem-cache-update}.

\subsubsection{Overview of the Barrier Mechanism}\label{subsubsec:barrier-overview}

In this proposal, Reads on the objects subject to load balancing are
blocked the moment a change in Virtual Node assignment is detected,
which prevents Reconcile from accessing a stale cache.

Only the objects of the newly assigned Virtual Nodes are then Listed
with a Label and added to the cache, so that the target of the List at
reassignment is limited to the objects of the new Virtual Nodes rather
than the total number of objects.
The block is released once the result of the List is reflected in the
cache.

This procedure guarantees that Reconcile always references a cache
that corresponds to the new assignment.

\subsubsection{Blocking Dependent Resources}\label{subsubsec:dependent-resources-barrier}

The Barrier mechanism blocks Gets not only on the resource that
received the assignment-change notification but also on the resources
that depend on it.
What a dependency between resources means, and why dependent resources
have to be blocked, is explained below.

That resource A depends on resource B means that the processing of an
object of A requires referencing an object of B.
The parent and child objects described in
Section~\ref{subsubsec:parent-child-same-pod} are a typical example:
since the child object is referenced while the parent object is
processed, the resource of the parent depends on the resource of the
child.

As described in Section~\ref{subsec:problem-cache-update}, the
processing of a change in Virtual Node assignment is carried out
independently for each resource, so that the timing of the processing
differs between resources and an inconsistency occurs because the
cache of a dependent resource remains stale.
Were the Barrier mechanism to block only the Gets of the resource that
received the assignment-change notification, the Gets of the dependent
resource would remain unblocked, and this inconsistency could not be
prevented.

To forestall this problem, the moment an assignment-change
notification is received for some resource, Gets are blocked not only
for that resource itself but also for every resource on which it
depends.
A Get on the child that is called during the Reconcile of the parent
is thereby blocked even before the assignment-change notification for
the resource of the child has arrived.
The block on each resource is released once the List for that resource
itself has completed.

\subsubsection{Handling Consecutive Reassignments with the Revision}\label{subsubsec:revision-sequential-changes}

As described in Section~\ref{subsec:problem-cache-update}, when the
addition or removal of instances occurs in a short period of time, the
next assignment change may arrive before the result of the first List
has been written into the cache in full.
Were the block released as soon as the first assignment change
completed, Reconcile would reference an incomplete cache that did not
reflect the second one.

To cope with this problem, every Virtual Node assignment-change event
carries, as its ID, the Revision at the moment the membership changed.
A Revision is a unique integer that increases monotonically across the
entire data store, and the same value is guaranteed on all replicas.

The Barrier mechanism uses this Revision to manage the start and the
release of the block.
Each resource holds, as ``\texttt{pendingRVs}'', the set of the
Revisions whose assignment change is being processed.
When an assignment-change notification is received, its Revision is
added to ``\texttt{pendingRVs}''; when a List completes, every entry
whose Revision is at most the completed one is removed from
``\texttt{pendingRVs}''.
Entries below the completed Revision are removed as well, and not
merely the completed Revision itself, in order to prevent past
Revisions from remaining in ``\texttt{pendingRVs}'' indefinitely
should a List or a Watch fail for some reason.
The block is in effect while ``\texttt{pendingRVs}'' is non-empty and
is released once the set becomes empty.

Algorithm~\ref{alg:barrier-pending} presents the processing on the
Barrier side as pseudocode.

\begin{algorithm}[t]
\caption{Block management in the Barrier mechanism via ``\texttt{pendingRVs}''}
\label{alg:barrier-pending}
\SetKwFunction{NotifyBarrier}{NotifyBarrier}
\SetKwFunction{NotifyBarrierComplete}{NotifyBarrierComplete}
\SetKwFunction{Get}{Get}
\SetKwProg{Fn}{Function}{:}{}

\Fn{\NotifyBarrier{$\mathit{resource}$, $\mathit{revision}$}}{
  $\mathit{pendingRVs}[\mathit{resource}] \leftarrow \mathit{pendingRVs}[\mathit{resource}] \cup \{\mathit{revision}\}$\;
  \ForEach{$\mathit{dep} \in \mathrm{Dependencies}(\mathit{resource})$}{
    $\mathit{pendingRVs}[\mathit{dep}] \leftarrow \mathit{pendingRVs}[\mathit{dep}] \cup \{\mathit{revision}\}$\;
  }
}
\BlankLine
\Fn{\NotifyBarrierComplete{$\mathit{resource}$, $\mathit{revision}$}}{
  \ForEach{$r \in \mathit{pendingRVs}[\mathit{resource}]$}{
    \If{$r \leq \mathit{revision}$}{
      $\mathit{pendingRVs}[\mathit{resource}] \leftarrow \mathit{pendingRVs}[\mathit{resource}] \setminus \{r\}$\;
    }
  }
}
\end{algorithm}

This mechanism also copes with consecutive assignment changes.
When, for example, an assignment change with Revision $R_1$ arrives,
$\mathit{pendingRVs} = \{R_1\}$ and the block starts.
If another change with a larger Revision $R_2$ arrives while the List
is in progress, $\mathit{pendingRVs} = \{R_1, R_2\}$.
When the List for $R_1$ completes, $R_1$ is removed and
$\mathit{pendingRVs} = \{R_2\}$, but the block continues because the
set is still non-empty.
When the List for $R_2$ completes, $R_2$ is removed as well,
$\mathit{pendingRVs} = \emptyset$, and the block is released at this
point.

Even when assignment changes occur consecutively, it is thus
guaranteed that Reconcile is always given a cache that corresponds to
the latest assignment.

\section{Implementation}\label{sec:implementation}

This section describes the implementation of the method proposed in
Section~\ref{sec:proposal}.
For this study, we forked Kubernetes and controller-runtime separately
and carried out the development on those forks.
The prerequisite knowledge needed to understand the implementation is
explained first, and the details of the implementation follow.

\subsection{Prerequisite Knowledge}\label{subsec:implementation-prerequired-knowledge}

A controller instance in the model of
Section~\ref{sec:kubernetes-model} corresponds to a Controller Pod, a
Pod being the unit of container execution in Kubernetes.
We used etcd~\cite{etcd} as the data store, which is one
implementation of the data store that Kubernetes adopts by default.
The filtering of objects through the Label mechanism is called a Label
Selector.

\begin{figure}[t]
  \centering
  \includegraphics[width=\columnwidth]{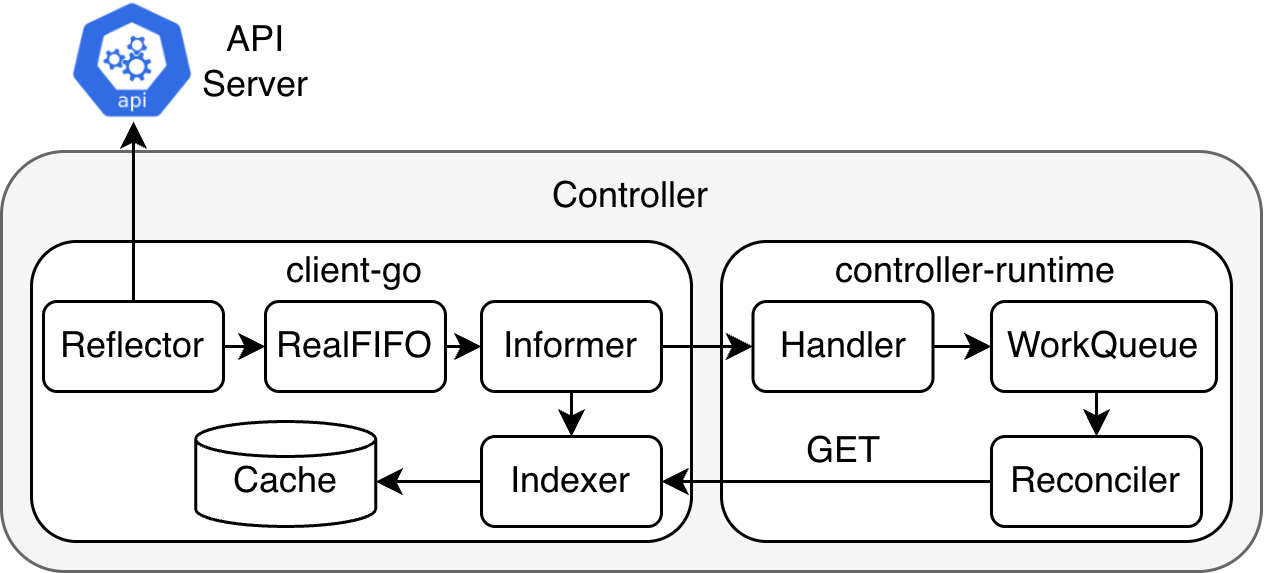}
  \Description{A block diagram showing the internal architecture of a Controller. The client-go section contains Reflector, RealFIFO, Informer, Indexer, and Cache connected in a pipeline. The Reflector receives data from the API Server. The controller-runtime section contains Handler, WorkQueue, and Reconciler. The Reconciler performs GET operations on the Indexer.}
  \caption{Operation of client-go and controller-runtime.}
  \label{fig:client-go-controller-runtime}
\end{figure}

Two libraries, client-go and controller-runtime, are widely used to
implement Kubernetes controllers, and
Figure~\ref{fig:client-go-controller-runtime} shows how they operate.

client-go~\cite{client-go} is the official Go client library for
interacting with the Kubernetes API.
It provides a mechanism that notifies changes to objects in real time
while minimizing the queries issued to the API Server.
Its main components are the Reflector, the RealFIFO, the Informer, and
the Indexer, one instance of each being created per resource.

The Reflector establishes a Watch connection with the API Server and
receives change events continuously.
The events it receives are passed to the RealFIFO and processed in
order.
When the same Controller Pod watches multiple resources, a separate
Reflector establishes a Watch connection with the API Server for each
resource.

The RealFIFO is a queue that conveys events from the Reflector to the
Informer.

The Informer dequeues events from the RealFIFO, notifies the Indexer
of the changes, and delivers the events to event handlers as well.

The Indexer is an in-memory cache that stores the latest state of the
objects, and it corresponds to the cache held by a controller instance
in the model.
A controller reads the state of an object from the Indexer rather than
querying the API Server directly, which eliminates redundant API calls
and reduces the load on the API Server.

controller-runtime~\cite{controller-runtime-github} is a framework
built on top of client-go that abstracts the configuration of
Informers and the management of the Reconcile loop.
A controller developer need only declare the objects to watch and
their relationships, for example by
``\texttt{For(Parent).Owns(Child)}'', and the Informers required are
configured automatically.
In this study, we build the dependency graph of resources from this
declaration.
In controller-runtime, a call to the Get or the List method reads from
the in-memory cache of client-go, and no access to the API Server
occurs.

\subsection{Implementation Target}\label{subsec:implementation-target}

We forked Kubernetes and controller-runtime separately for
development, and the fork of Kubernetes includes both the API Server
and client-go.

Kubernetes~\cite{kubernetes-github} was forked from commit
``\texttt{6e753bd}'', and
controller-runtime~\cite{controller-runtime-github} was forked from
commit ``\texttt{6210f84}''.

client-go has feature flags called Feature Gates, which toggle new
features on and off.
Table~\ref{tab:feature-gates} shows the Feature Gate settings used in
the evaluation environment of this study, and the purpose of each
Feature Gate is described in the relevant section.

\begin{table}[t]
  \caption{client-go Feature Gate settings.}
  \label{tab:feature-gates}
  \centering
  \begin{tabular}{l|c}
    \hline
    Feature Gate & State \\
    \hline
    ``\texttt{AtomicFIFO}'' & true \\
    ``\texttt{InOrderInformers}'' & true \\
    \hline
  \end{tabular}
\end{table}

\subsection{API Server Implementation}\label{subsec:api-server-implementation}

\subsubsection{Choice of Hash Function}\label{subsubsec:hash-function-selection}

We adopted CRC32 as the hash function.
CRC32 is deterministic, that is, its output for a given input string
is always the same, regardless of the environment or the process.
This determinism guarantees that the same hash value is computed for
the same input even on a different process or a different machine.

In the second-stage Consistent Hash Ring, moreover, each instance is
placed on the ring as multiple replicas, an approach inspired by
DeCandia et al.~\cite{decandia2007}.
A uniform load distribution is thereby achieved even when the number
of instances is small.

\subsubsection{Opt-In of the Virtual Node Label and the Hash Key}\label{subsubsec:virtual-node-label-opt-in}

We implemented the system in such a way that the objects subject to
load balancing opt in explicitly through an Annotation.
There are two ways to opt in.
The first is to set the ``\path{loadbalance.k8s.io/loadbalanced}''
Annotation to ``\texttt{true}''.
The second is to set the ``\path{loadbalance.k8s.io/hash-key}''
Annotation to an arbitrary Hash Key.
The latter serves both as the opt-in and as the means of specifying
the Hash Key that assigns parent and child objects to the same Pod, as
described in Section~\ref{sec:proposal}.

The attachment of the Virtual Node Label to an object is implemented
as a Mutating Admission Webhook as a proof of concept.
When a request to create or to update an object reaches the API
Server, the webhook performs the opt-in check and attaches the Virtual
Node number computed from the first-stage modular hash to the object
as the ``\texttt{loadbalance.k8s.io/vn}'' Label.

The number of Virtual Nodes is fixed at 200.
This value is determined before the system starts and is not changed
during operation.

\subsubsection{Virtual Node Assignment Delivery Endpoint}\label{subsubsec:resource-assignment-endpoint}

To deliver Virtual Node assignments to each Controller Pod, we added a
``\texttt{/virtual\allowbreak node\allowbreak slices}'' endpoint to the API Server.
A Controller Pod connects to this endpoint over HTTP, specifying the
Watch Group name and the memberID as query parameters.

\begin{verbatim}
GET /virtualnodeslices
    ?group=parent-controllers
    &member_id=pod-a
    &watch=true
\end{verbatim}

The API Server delivers events as a JSON stream.
Each event contains the memberID, the etcd Revision, and the array of
the Virtual Node numbers assigned to that instance.

\begin{verbatim}
{
  "type": "REBALANCE",
  "object": {
    "apiVersion":
        "loadbalance.k8s.io/v1",
    "kind": "VirtualNodeSlice",
    "metadata": {
      "name": "pod-a",
      "resourceVersion": "12345"
    },
    "spec": {
      "group": "parent-controllers",
      "memberID": "pod-a",
      "vns": [0, 5, 10, 15, 20]
    }
  }
}
\end{verbatim}

The ``\texttt{metadata.resourceVersion}'' field stores the etcd
Revision.
This value is used as the ID for the start and the release of the
block in the Barrier mechanism described in
Section~\ref{sec:proposal}.

\subsubsection{Storing Member Entries in etcd}\label{subsubsec:member-entry-etcd-storage}

A member entry is stored in etcd under the key
``\path{/watchgroups/{groupName}/{memberID}}''.
If, for example, the Watch Group name is
``\texttt{parent-controllers}'' and the memberID is
``\texttt{pod-a}'', the key is
``\path{/watchgroups/parent-controllers/pod-a}''.

A member entry is held as a single cluster-wide registration rather
than one registration per resource.
This design was chosen because the Barrier mechanism uses the etcd
Revision to manage the block across all resources collectively: since
a change in membership is notified with the same etcd Revision
regardless of the resource, the Reflectors of different resources can
List and release their blocks on the basis of the same Revision.

A member entry is written together with an etcd Lease~\cite{k8s-leases}
whose TTL is 30 seconds.
Under normal operation, the API Server revokes the Lease as soon as
the Watch connection is broken, without waiting for the 30-second TTL
to elapse.
Should the API Server fail to perform the revocation for some reason,
etcd nevertheless deletes the entry automatically once the TTL has
elapsed.

\subsection{client-go Implementation}\label{subsec:client-go-implementation}

\subsubsection{Reflector Implementation}\label{subsubsec:reflector-virtual-node-reception}

We added to the Reflector of client-go a mechanism that watches the
``\texttt{/virtual\allowbreak node\allowbreak slices}'' endpoint.
Each Reflector receives its own Virtual Node assignment from the API
Server and attaches a Label Selector based on the Virtual Node numbers
to its List and Watch requests.
If, for example, the Virtual Nodes 0, 5, and 10 are assigned, the
Label Selector ``\texttt{loadbalance.k8s.io/vn in (0,5,10)}'' is
specified.

To convey the result of the processing of an assignment change, we
defined a new ``\texttt{Rebalanced}'' event dedicated to rebalancing.
Table~\ref{tab:rebalanced-event} shows the properties that the
``\texttt{Rebalanced}'' event carries.
The Reflector delivers this event to the Informer through the
RealFIFO.

\begin{table}[t]
  \caption{Properties of the ``\texttt{Rebalanced}'' event.}
  \label{tab:rebalanced-event}
  \centering
  \begin{tabular}{l|p{5cm}}
    \hline
    Property & Meaning \\
    \hline
    ``\texttt{ResourceVersion}'' & etcd Revision at the time of the membership update \\
    ``\texttt{AddedObjects}'' & Objects newly retrieved by List \\
    ``\texttt{VNs}'' & List of Virtual Node assignments \\
    \hline
  \end{tabular}
\end{table}

The Reflector keeps the set of the Virtual Node numbers for which the
List has completed, and it performs the following procedure when it
receives a notification of a change in Virtual Node assignment.
\begin{enumerate}
  \item Notify controller-runtime that a change in Virtual Node
        assignment has been received.
  \item Stop the current Watch.
  \item Compare the Virtual Nodes listed previously with the current
        Virtual Node assignment to identify the newly added Virtual
        Nodes.
  \item List, with a Label Selector, only the objects of the newly
        added Virtual Nodes.
  \item Add a ``\texttt{Rebalanced}'' event to the queue.
  \item Resume the Watch with the new Label Selector.
\end{enumerate}
If the List fails, the procedure starts over from the initial List.

As mentioned in
Section~\ref{subsec:implementation-prerequired-knowledge}, one
Reflector is created per resource.
Owing to implementation constraints, when the same controller watches
multiple resources, a separate Reflector for each resource watches the
same ``\texttt{/virtual\allowbreak node\allowbreak slices}'' endpoint.

\subsubsection{Informer Implementation}\label{subsubsec:rebalanced-event-notification}

Upon receiving a ``\texttt{Re\-balanced}'' event, the Informer performs
the following processing.
\begin{enumerate}
  \item Convey the new list of Virtual Nodes to the Indexer.
  \item Add the newly retrieved objects to the Indexer.
  \item Remove the objects of the Virtual Nodes that are no longer in
        the assignment.
  \item Update the ``\texttt{resourceVersion}'' recorded in the
        Indexer.
  \item Notify the Barrier mechanism of controller-runtime of the etcd
        Revision to signal that the List has completed.
\end{enumerate}
The Barrier mechanism uses this Revision to decide when to release the
block.

\subsubsection{Managing the Virtual Node List in the Indexer}\label{subsubsec:indexer-vn-storage}

We added dedicated storage to the Indexer for managing the list of
Virtual Nodes.

\subsection{controller-runtime Implementation}\label{subsec:controller-runtime-implementation}

As mentioned in
Section~\ref{subsec:implementation-prerequired-knowledge}, the
relationships among objects are declared in controller-runtime when a
controller is constructed, for example by
``\texttt{For(Parent).Owns(Child)}''.
We built a dependency graph of resources from this declaration.

For the Barrier mechanism, we implemented a Barrier Manager in
controller-runtime and managed the set of pending Revisions for each
resource.

In addition, we added to controller-runtime a check, performed before
Reconcile is invoked, that verifies whether the Virtual Node of the
object is assigned to the instance itself.

\section{Evaluation}\label{sec:evaluation}

This section describes the evaluation of the proposed method.

\subsection{Evaluation Environment}\label{subsec:evaluation-environment}

The environment used for the evaluation experiments is described
first.
The experiments were run on the physical machines shown in
Table~\ref{tab:physical-machines}.

\begin{table}[t]
  \caption{Physical machines used.}
  \label{tab:physical-machines}
  \centering
  \footnotesize
  \begin{tabular}{l|l}
    \hline
    Machine & CPU \\
    \hline
    ua & Intel Xeon Silver 4116 (12C/24T $\times$2) \\
    lh & AMD EPYC 7452 (32C/64T) \\
    gz & Intel(R) Xeon(R) 6746E  \\
    \hline
  \end{tabular}
\end{table}

On these physical machines we built virtual machines and a Kubernetes
cluster with the configuration shown in
Table~\ref{tab:vm-configuration}.
Only the controller under evaluation was placed on worker8, and the
other components were placed on worker1 through worker7.

\begin{table}[t]
  \caption{VM configuration.}
  \label{tab:vm-configuration}
  \centering
  \small
  \begin{tabular}{l|r|c|r|r}
    \hline
    Node & Count & Host & vCPU & Memory \\
    \hline
    etcd & 1 & ua & 8 & 16GB \\
    master1--3 & 3 & ua & 8 & 16GB \\
    worker1--3 & 3 & ua & 12 & 32GB \\
    worker5--7 & 3 & lh & 12 & 32GB \\
    worker8 & 1 & gz & 100 & 32GB \\
    \hline
  \end{tabular}
\end{table}

\subsection{Basic Evaluation}\label{subsec:basic-evaluation}

In the basic evaluation, we assessed the performance of the proposed
method with a simple Kubernetes controller built for the purpose.
The components used in the evaluation are described first, and two
experiments and their results follow.

\subsubsection{Evaluation Components}\label{subsubsec:evaluation-components}

The basic evaluation uses the following three components.
\begin{itemize}
  \item Sample Controller
  \item Experiment Operator
  \item Parent Scraper
\end{itemize}

The Sample Controller is the controller under evaluation.
It is a simple Kubernetes controller that watches Parent objects and
creates one corresponding Child object during the Reconcile of each
Parent object, and it runs load balanced by the proposed method.
A Gate object makes it possible to stop and to resume the creation of
Child objects at any time.
So that the load of the other components should not affect the
measurements, the Sample Controller alone was placed on worker8.

The Experiment Operator is an operator that accepts an object of the
ParentCreation custom resource and creates the specified number of
Parent objects at once.
It was placed on worker1 through worker7.

The Parent Scraper is a component that monitors the state of Child
object creation for all Parent objects.
It exposes as Prometheus metrics the total number of Parent objects
and the number of Parent objects whose corresponding Child object has
been created, and it serves to determine when all Child objects have
been created.
For load balancing, it ran distributed over five replicas by the
proposed method.
It was placed on worker1 through worker7.

\subsubsection{Experiment 1: Relationship between Pod Count and Reconcile Throughput}\label{subsubsec:experiment-reconcile-speed}

We first measured how the Reconcile throughput changes as the number
of Sample Controller Pods varies.
The cluster was built with the number of Virtual Nodes set to
1{,}000.

The internal concurrency of each Pod was set to 5, which means that
five workers, as described in Section~\ref{sec:kubernetes-model}, run
on each Pod.
The concurrency was set to 5 because many standard Kubernetes
controllers use a default concurrency of 5.

The scenario of the experiment is as follows.
\begin{enumerate}
  \item Set the number of Sample Controller replicas to a specified
        value.
  \item Instruct the Experiment Operator to create 100{,}000 Parent
        objects.
  \item Wait for the creation of the Parent objects to complete while
        the Gate is closed.
  \item Open the Gate to start the processing of the Sample
        Controller.
  \item Measure the time until the Child objects for all the Parent
        objects have been created.
\end{enumerate}
We varied the number of Sample Controller replicas from 1 to 9 and
performed three runs for each setting.

\begin{figure}[t]
  \centering
  \includegraphics[width=\columnwidth]{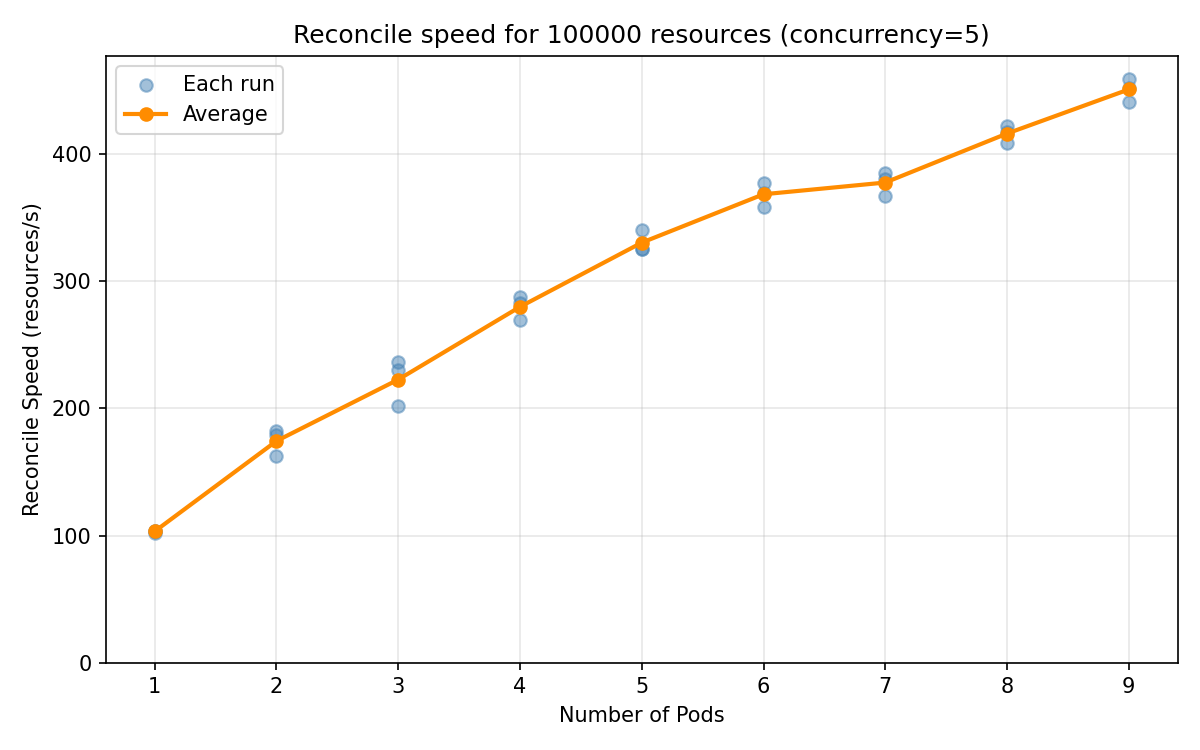}
  \Description{A line chart with Number of Pods on the horizontal axis (1 to 9) and Reconcile Speed in resources per second on the vertical axis (0 to 450). Individual run data points are shown as light blue dots and the average is shown as an orange line. The average speed increases from about 100 at 1 Pod to about 450 at 9 Pods.}
  \caption{Relationship between Pod count and Reconcile throughput.}
  \label{fig:reconcile-speed}
\end{figure}

Figure~\ref{fig:reconcile-speed} shows the result.
The horizontal axis is the number of Sample Controller Pods, and the
vertical axis is the Reconcile throughput in objects per second.
The Reconcile throughput improves as the Pod count grows, which
confirms that the horizontal scaling by the proposed method works
effectively.

The variation in performance across the runs is due to the
characteristics of the Consistent Hash Ring.

\subsubsection{Experiment 2: Relationship between the Number of Virtual Nodes and the Reassignment Duration}\label{subsubsec:experiment-barrier-duration}

Next, we measured the time required, when a member entry changes, from
the change in membership to the completion of the block of the Barrier
mechanism.

The scenario of the experiment is as follows.
\begin{enumerate}
  \item Set the number of Sample Controller replicas to 1.
  \item Create 10{,}000 Parent objects and wait for all the Child
        objects to be created.
  \item Repeat the registration and the deregistration of the member
        entry 100 times at 30-second intervals.
  \item For each switch, measure the reassignment duration, that is,
        the time from the change in membership to the end of the
        Barrier block, and take the average over the 100 trials.
\end{enumerate}
The experiment was carried out with the number of Virtual Nodes varied
over 100, 1{,}000, 10{,}000, and 100{,}000.

\begin{figure}[t]
  \centering
  \includegraphics[width=\columnwidth]{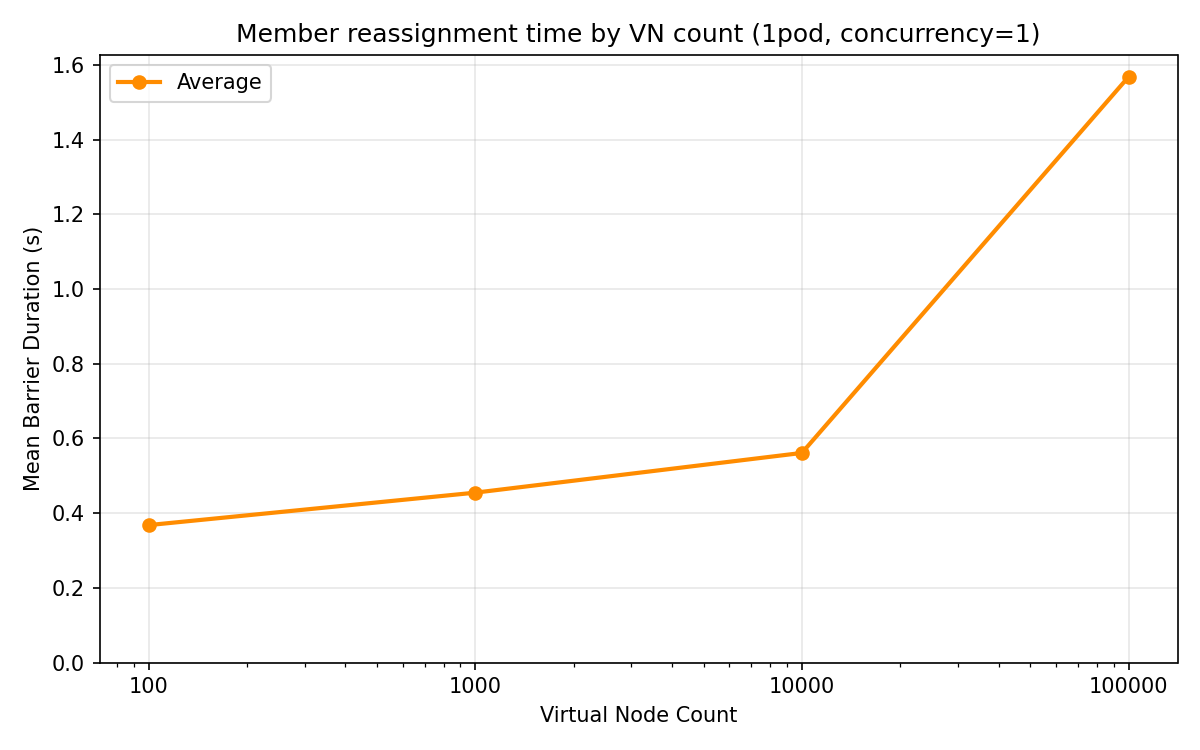}
  \Description{A line chart with Virtual Node Count on the horizontal axis (logarithmic scale, from 100 to 100000) and Mean Barrier Duration in seconds on the vertical axis (0 to 1.6). The orange line shows the average duration stays around 0.37 to 0.56 seconds for VN counts of 100 to 10000, then rises sharply to about 1.57 seconds at 100000.}
  \caption{Relationship between the number of Virtual Nodes and the reassignment duration. The number of objects is 10{,}000.}
  \label{fig:barrier-duration}
\end{figure}

Figure~\ref{fig:barrier-duration} shows the result.
The horizontal axis is the number of Virtual Nodes on a logarithmic
scale, and the vertical axis is the average reassignment duration in
seconds.
The duration tends to increase once the number of Virtual Nodes
reaches 10{,}000 or more, but it stays around 0.5 seconds when the
number of Virtual Nodes is between 100 and 1{,}000.
Since the Reconcile of a Kubernetes controller does not demand
immediate responsiveness, this duration is within an acceptable range.

\section{Discussion}\label{sec:discussion}

This section clarifies what the proposed method can and cannot
achieve.

In the proposed method, accesses to the cache are blocked from the
moment a reassignment occurs until the cache update has completed,
which eliminates the problem of accessing a stale cache that
conventional approaches suffer from.

The following problems, on the other hand, are not addressed by this
study.

The first problem is that the same object can momentarily be assigned
to several instances during a reassignment.
The moments at which different instances receive the
assignment-change notification differ by a few milliseconds.
If one instance completes the reassignment before another instance has
received the notification, the same object is temporarily managed by
several instances, and the cluster may temporarily enter an unintended
state if Reconcile is called for that object at the same time.
Controller developers need to write their programs with this in mind.
Since the Kubernetes control plane assumes eventual consistency and is
designed so that convergence to the desired state over time suffices,
entering an unintended state temporarily is not a major problem.

The second problem is that a Get on an object that is not assigned to
the controller does not return the correct result.
This problem is common to the existing approaches as
well~\cite{fluxcd,externaldns,cert-manager,ingress-nginx,prometheus-operator,argocd,kubevela,knative,kep5866,ebert2024}.
In the proposed method, an object that is not assigned to the instance
itself does not enter the cache, while an instance consults only its
own cache on a Get and sends no request to the API Server.
The problem arises for these reasons.
Controller developers therefore need to write their programs with
care, so that the objects an instance needs to reference are assigned
to it correctly.

\section{Conclusion}\label{sec:conclusion}

In this study, we solved the two problems from which Ebert's
method~\cite{ebert2024} for the horizontal scaling of the Kubernetes
control plane suffered.
The first problem was that the reassignment of objects incurred Label
rewrites on the API Server and the data store in proportion to the
number of affected objects.
For this problem, we introduced object assignment based on a two-level
Hash.
By designing the Label of an object to hold its Virtual Node number
rather than the information about the responsible instance, and by
keeping the first-stage mapping constant throughout the lifetime of
the system, we eliminated Label rewrites on objects entirely when
instances are added or removed.
The second problem was that, because Watch and Reconcile are performed
in parallel for each resource, accesses to a stale cache could occur
when objects were reassigned.
For this problem, we introduced the Barrier mechanism, which, the
moment a reassignment is detected, temporarily blocks cache accesses
to the resource concerned and to the resources that depend on it,
Lists only the objects of the newly assigned Virtual Nodes and adds
them to the cache, and then releases the block.
Reconcile is thereby guaranteed always to reference a cache that
corresponds to the new assignment.

We also forked Kubernetes itself and controller-runtime, and
implemented the proposed method.
In the basic evaluation, we conducted two experiments with a Sample
Controller built for the purpose.
The first experiment confirmed that the Reconcile throughput improves
as the number of instances increases.
The second confirmed that the reassignment duration stays around 0.5
seconds as long as the number of Virtual Nodes is between 100 and
1{,}000.

As a direction for future work, we plan to propose this method
officially as a Kubernetes Enhancement Proposal (KEP).
Since the method requires modifications to Kubernetes itself and to
controller-runtime, it is desirable for the feature to be built into
Kubernetes by default so that every controller developer can use it
easily.
We will therefore propose this method as a KEP and aim for it to
become a standard feature of Kubernetes.

\begin{acks}
Claude codes was utilized to generate codes and texts based on the authors' ideas with interactive feedback with the authors. Final version of this paper was reviewed and edited by the authors.
\end{acks}

\bibliographystyle{ACM-Reference-Format}
\bibliography{related-works}

\end{document}